# Likelihood Analysis of Large-Scale Flows

Andrew H. Jaffe and Nick Kaiser

*Canadian Institute for Theoretical Astrophysics,*
*60 St. George St., Toronto, Ontario M5S 1A1, Canada*

**ABSTRACT**

We apply a likelihood analysis to the data of Lauer & Postman 1994 With $P(k)$ parametrized by $(\sigma_8, \Gamma)$, the likelihood function peaks at $\sigma_8 \simeq 0.3$, $\Gamma \lesssim 0.025$, indicating at face value very strong large-scale power, though at a level incompatible with COBE. There is, however, a ridge of likelihood such that more conventional power spectra do not seem strongly disfavored. The likelihood calculated using as data only the components of the bulk flow solution peaks at higher $\sigma_8$, in agreement with other analyses, but is rather broad. The likelihood incorporating both bulk flow and shear gives a different picture. The components of the shear are all low, and this pulls the peak to lower amplitudes as a compromise.

The Lauer & Postman velocity data alone are therefore *consistent* with models with very strong large scale power which generates a large bulk flow, but the small shear (which also probes fairly large scales) requires that the power would have to be at *very* large scales, which is strongly disfavored by COBE. The velocity data also seem compatible with more conventional $P(k)$ with $0.2 \lesssim \Gamma \lesssim 0.5$, and the likelihood is peaked around $\sigma_8 \sim 1$, in which case the bulk flow is a moderate, but not extreme, statistical fluctuation.

Applying the same techniques to the data of Riess, Press, & Kirshner 1995, the results are quite different. The flow is not inconsistent with the microwave dipole and we derive only an upper limit to the amplitude of the power spectrum: $\sigma_8 \lesssim 1.5$ at roughly 99%.

*Subject headings:* Cosmology: theory – observation – galaxies: clustering – distances and redshifts

## 1. Introduction

There have recently been several analyses of the large-scale velocity data for 119 Abell clusters within 15000 km/s obtained by Lauer and Postman (1994, hereafter LP) which suggest that most currently popular models of structure formation (*e.g.*, CDM and its variants obtained by adding tilt, a cosmological constant, or an admixture of hot dark matter) could not produce the large magnitude of the bulk flow seen, at the "2–3$\sigma$ level" (Tegmark, Bunn & Hu 1993, Strauss *et al.* 1994, Feldman & Watkins 1994). All of these have focused on the bulk flow statistic, a highly reduced representation of the data. Here, we will perform a likelihood analysis which uses more of the data. We take as our "hypothesis space" the family of power spectra $P(k;\sigma_8,\Gamma)$, parametrized by the usual amplitude and shape parameters $\sigma_8$ and $\Gamma$ (Efstathiou, Bond & White 1992; see Eq. (7)), and calculate the likelihood $\mathcal{L}(\sigma_8,\Gamma) = P(\text{data}|\sigma_8\Gamma)$. We also calculate the likelihood for the case when the data are reduced to either the bulk flow alone (as in other analyses) or the bulk flow and a gradient. Since the gradient or shear also probes large scales, we would expect these to give results compatible with the bulk flow alone.

More recently, Riess, Press & Kirshner (1995, hereafter RPK) have analyzed data from a nascent survey employing the light curves of Supernovae Ia as distance indicators; although the sample of 13 galaxies out to 7000 km/s is quite sparse, the errors ($\approx 5\%$) are small enough to make the data interesting. We apply our techniques to this dataset as well.

## 2. The Likelihood Function

The main statistical tool we will use to analyze the peculiar velocity data will be Bayes' Theorem, and through it, the likelihood function, as in Kaiser 1988. Bayes' theorem can be written as

$$p(\theta|DI) = \frac{p(\theta|I)p(D|\theta I)}{p(D|I)} \qquad (1)$$

where $p(a|bc)$ roughly means "the probability [density] of $a$ given $b$ and $c$." Here, $\theta$ represents the parameters of the theory we are considering (here, $\sigma_8$ and $\Gamma$), $D$ represents the data, and $I$ represents any prior information we bring to the problem. Thus, $p(\theta|I)$ is the infamous "prior distribution" of the parameters, $p(D|\theta I) = \mathcal{L}(\theta; D)$ is the likelihood function, and the remaining factor in the denominator merely serves as a normalizing constant so that $\int d\theta\, p(\theta|DI) = 1$. Eq. (1) tells us how we update the probabilities one assigns to various hypothetical power spectra. Other information (*e.g.*, the normalization from COBE—although we do not use this information in this paper) can go into the prior.

The data are given by a list of particle positions, velocities, and errors, calculated from the data. LP use the brightest galaxy in a cluster as a distance indicator, with a phenomenological relationship between the absolute magnitude, $L_m$, and the slope of the brightness profile of the galaxy, $\alpha$. RPK use the shape of the SNIa light curve as a predictor of its luminosity.

We shall assume that the *initial* density and velocity fields had a Gaussian distribution with some power spectrum $P(k) = \langle|\delta_k|^2\rangle$. Assuming linear evolution, the velocity power spectrum is

$$P_v(k) = \langle|v_k|^2\rangle = \left(\frac{Ha}{k}\right)^2 P(k) \qquad (2)$$

for a universe with critical density; for $\Omega \neq 1$, this would be modified by an additional factor of $f(\Omega)^2 \approx \Omega^{1.2}$. Under the assumption of Gaussian initial conditions, these power spectra provide a complete statistical description of the density and velocity fields. We assume that the errors are Gaussian, as well. We can therefore write the likelihood function as

$$\mathcal{L} = \frac{1}{(2\pi)^{N/2}|R|^{1/2}} \exp\left(-\frac{1}{2}S_m R_{mn}^{-1} S_n\right) \qquad (3)$$

where $S_m$ are the line-of-sight peculiar velocities for the $m = 1$ to $N$ data points, and $R_{mn} = \langle S_m S_n \rangle$ defines the correlation matrix for the data. We can split this into two independent terms: a "theoretical" covariance matrix $R^{(v)}$ and an "error" matrix $R^{(e)}$.

If the galaxies are located at positions $\mathbf{r}_m = \hat{\mathbf{r}}_m r_m$, the theoretical covariance matrix is given by

$$\begin{aligned} R^{(v)}_{mn} &= \langle \hat{\mathbf{r}}_m \cdot \mathbf{v}_m \hat{\mathbf{r}}_n \cdot \mathbf{v}_n \rangle = \hat{r}_{m,i}\hat{r}_{n,j}\langle v_{m,i} v_{n,j}\rangle \\ &= \int \frac{4\pi k^2 dk}{(2\pi)^3} P_v(k) f_{mn}(k). \end{aligned} \qquad (4)$$

where we sum over the spatial indices $i,j$, but not over the cluster indices $m,n$, and the function $f_{mn}(k)$ is the angle-average given by

$$f_{mn}(k) = \hat{r}_{m,i}\hat{r}_{n,j}\int \frac{d^2\hat{k}}{4\pi} \hat{k}_i \hat{k}_j e^{i\mathbf{k}\cdot(\mathbf{r}_m - \mathbf{r}_n)}. \qquad (5)$$

The other part of the correlation matrix simply adds in the velocity errors:

$$R^{(e)}_{mn} = (\sigma_m^2 + \sigma_*^2)\delta_{mn} \qquad \text{no sum} \qquad (6)$$

where $\sigma_m$ is the velocity error for cluster $m$. The term $\sigma_*$ is added to take into account a variety of effects: LP report a 1-d dispersion of 271 km/s between the BCG redshift and the average cluster redshift; we also must take into account the possibility that the velocity field has undergone some nolinear evolution. For field galaxies, this effect may be large, but for clusters it is expected to be small. In total, we choose $\sigma_* = 350$ km/s for the results we present; however the results are not strongly dependent on this value—the approximately $.16cz$ errors in the BCG velocity are dominant; for the SNIa, even the 5% distance indicator errors still dominate, although less so. We note in passing that we have not taken into account the errors in the *positions* of the clusters (which are equal to those in the velocities, since $cz = H_0 r + \hat{\mathbf{r}} \cdot \mathbf{v}$); for the LP data, we do, however, use the expressions from Colless (1995) to calculate the "Hubble redshift" $cz_H \approx cz - S$ that the galaxy would have in the absence of peculiar motions, which is in turn used to calculate the comoving distance to the cluster $d \approx cz_H/H_0$. We are also using Colless' fit to the $L_m - \alpha$ relation, which assumes a random distribution of velocities aside from the bulk flow.

We model the spectrum as an initial ($n = 1$ Harrison-Zel'dovich) power law times an appropriate transfer function: $P(k) \propto k T^2(k)$ We choose the following ansatz, after Efstathiou, Bond & White 1992:

$$T(k) = \left[1 + \left(ak/\Gamma + (bk/\Gamma)^{3/2} + (ck/\Gamma)^2\right)^\nu\right]^{-1/\nu} \quad (7)$$

with $a = 6.4h^{-1}$ Mpc, $b = 3.0h^{-1}$ Mpc, $c = 1.7h^{-1}$ Mpc, $\nu = 1.13$, leaving two free parameters, the overall amplitude of the spectrum, given for example by $\sigma_8$, and the value of $\Gamma$. The latter controls the location of the turnover from the large-scale power law $P(k) \propto k$ to the small-scale $P(k) \propto k^{-3}$. For a universe with cold dark matter, $\Gamma \approx \Omega h$, and $\Gamma$ can be also be related to quantities in universes with decaying neutrinos or mixed dark matter.

We have checked the method with realizations of Gaussian density and velocity fields, related by linear gravitational evolution. We do not simulate a particular distance indicator relation, but add unbiased Gaussian errors on the velocity realization, of comparable magnitude to those in the LP sample. We are able to recover the value of $\sigma_8$ to better than "one sigma" (*i.e.*, 68%), but $\Gamma$ is less tightly constrained (as it is for the actual data below).

In Fig. 1, we show the results of applying the method to the LP data. In order to help remove any possible nonlinear signal in the data, we have grouped the data using a friends-of-friends procedure with a linking scale of 2500 km/s. We perform an unweighted average of the clumps, assuming uncorrelated gaussian errors, leaving about 50 clumps of galaxies. Recall that throughout we use $\Omega = 1$; in general the results apply to the combination $f(\Omega)\sigma_8 \approx \Omega^{0.6}/b_8$, the usual $\beta$ parameter. The maximum likelihood value for the parameters is $\sigma_8 \simeq 0.3$, $\Gamma \simeq 0.025$; the data actually prefer low amplitudes for $\sigma_8$ with a peak in the power spectrum at very large scales. At this point, we have $\chi^2 \equiv S_m S_n R_{mn}^{-1} = 47.5$ for 53 degrees of freedom, the number of clumped clusters.

We show contours of constant likelihood, with likelihood ratios relative to maximum of 0.250, 0.201, 0.140, 0.079, and 0.019. Assuming a constant prior for $p(\sigma_8|I)$ and $p(\Gamma|I)$ over the plotted area, these likelihood ratios are such that the contours enclose 50%, 68%, 90%, 95%, and 99% of the accumulated posterior probability. (In addition, we show the 68% confidence limits for the COBE quadrupole $Q_{rms-PS} = 17^{+8}_{-5}$ $\mu K$ from Gorski *et al.* 1994, which can also be expressed as a normalization for the power spectrum amplitude, calculated on a much larger scale than $\sigma_8$, or the effective window function $\mathcal{W}^2(k)$.)

Preference for a low $\Gamma$ might have been anticipated from other analyses of the bulk flow, which seem to require a large amount of large-scale power. There is, however, a ridge in likelihood along increasing $\Gamma$. Canonical CDM, with $\sigma_8 = 1, \Gamma = 0.5$, has a likelihood ratio of 0.311, placing it within the 50% contour. A model with $\sigma_8 = 1, \Gamma = 0.2$ has a likelihood ratio of 0.376, also within the 50% contour. Note that we have assumed Gaussian errors; if the actual distribution has more power in the tails, then any limits may be considerably weaker than those quoted here.

[If we do not perform the clumping procedure, the results are even more restrictive, compressing the contours toward the $\sigma_8 = 0$ axis at all values of $\Gamma$. At $\Gamma \simeq 0.2$, the unclumped data require $\sigma_8 \lesssim 1.6$ at the 99% contour. In essence, this is an upper limit on detected power; however, we expect these results to be dominated by close associations of clusters where our linear results may not be applicable (Croft & Efstathiou 1994).]

Because LP (or Colless 1995) use the data itself to calibrate their BCG distance indicator, assuming that the flow is modelled by *only* a bulk flow, we must in-vestigate possible biases induced in using the individual peculiar velocities derived as residuals from BCG $L_m - \alpha$ relation. We have done this by using the calibration derived from both the best fit bulk flow and that assuming the CMB rest frame; the results are insensitive to this change—the mean $L_m - \alpha$ relation does not change substantially no matter what flow model is chosen. This applies to both the shapes of the likelihood contours and the values for the bulk flow and shear derived below.

Now, we turn to the RPK data; with only thirteen points within 7000 km/s, there is not much leverage on the shape of the power spectrum (*i.e.*, $\Gamma$), so we have simply set $\Gamma = 0.5$ and concentrate on $\sigma_8$. In Figure 2, we see that the RPK data in fact give an *upper limit* on $\sigma_8$; there is no clear detection of power, only consistency with moderate amplitudes for the power spectrum.

It is perhaps surprising that the RPK data, with only 13 galaxies, can put such a strong constraint on the amplitude of the power spectrum; this can be traced to the small error ($\approx 5\%$) in the SNIa distance indicator, a factor of three better than that of LP, somewhat making up for the factor of ten in sample size.

## 3. Bulk Flow and Shear Likelihoods

Why is the evidence against CDM-like models so different from conclusions drawn from analyses of the bulk flow in the LP data? To answer this, we now perform a likelihood analysis with reduced descriptions of the data, incorporating bulk flow and shear. In addition to allowing a comparison with other results, we expect the bulk flow and shear to be less affected by nonlinear evolution than the full cluster/galaxy velocity field (Croft & Efstathiou 1994, Bahcall, Cen & Grammann 1994).

We model the peculiar velocity field as $v_i(\mathbf{r}) = u_i + r_j p_{ij} + \cdots$ where $u_i$ is the bulk flow, $p_{ij}$ the shear tensor, and the series could obviously be extended to higher moments as desired. Alternatively, we could model the full velocity field including the Hubble expansion by the same bulk flow and an anisotropic expansion term, $u_i + r_j H_{ij} + \cdots$ with a "Hubble tensor" $H_{ij} = H_0 \delta_{ij} + \tilde{p}_{ij}$. However, we retain the trace of $p_{ij}$, because our estimators are different than those used by LP, and so we would not predict quite the same Hubble constant (although the difference is small; see Table 1).

Of course, we only have line-of-sight velocities, so we really only model $S(\mathbf{r}) = \hat{r}_i v_i(\mathbf{r})$ and we are only sensitive to the symmetric part of the shear tensor, as expected. Nonlinearity and observational error can be modeled as an additional error term, which we will take to be a gaussian.

The maximum likelihood values for $u_i$ and $p_{ij}$ (and any higher moments desired) can be considered as members of a 9-component vector $a_\alpha$; we are linearly fitting the $S(\mathbf{r}_n)$ to the nine independent functions $\hat{r}_i$, $\hat{r}_i \hat{r}_j r$, which we shall write as $g_p(\mathbf{r})$, $p = 1 \ldots 9$, with coefficients $a_p$. The maximum likelihood values for the $a_p$ are given by the usual linear fit for gaussian errors $\sigma_p$:

$$a_p = A_{pq}^{-1} \sum_n \frac{S_n g_q(\mathbf{r}_n)}{\sigma_n^2}; \quad A_{pq} = \sum_n \frac{g_p(\mathbf{r}_n) g_q(\mathbf{r}_n)}{\sigma_n^2} \quad (8)$$

(Lynden-Bell *et al.* 1988, Kaiser 1991). This expression can be converted to the integral

$$a_p = \int d^3 r \, W_{ip}(\mathbf{r}) v_i(\mathbf{r});$$
$$W_{ip}(\mathbf{r}) = A_{pq}^{-1} \sum_n \delta^3(\mathbf{r} - \mathbf{r}_n) \frac{\hat{r}_i g_p(\mathbf{r})}{\sigma(\mathbf{r})^2}. \quad (9)$$

(Restricted to $p = 1 \ldots 3$ so $g_p = \hat{r}_p$, this is equivalent to the expression in Kaiser 1988) Because the tensor window function is a linear filter on the field $v_i(\mathbf{r})$, the vector $a_p$ obeys the same Gaussian distribution, with a suitably altered correlation matrix,

$$\Sigma_{pq} = \langle a_p a_q \rangle = \int \frac{4\pi k^2 dk}{(2\pi)^3} P_v(k) \mathcal{W}_{pq}^2(k) + A_{pq}^{-1} \quad (10)$$

with an angle-averaged squared window function given by

$$\mathcal{W}_{pq}^2(k) = \int \frac{d^2 \hat{k}}{4\pi} \widetilde{W}_{ip}(\mathbf{k}) \widetilde{W}_{jq}(\mathbf{k}) \hat{k}_i \hat{k}_j \quad (11)$$

First performing this calculation for the three component (bulk flow only) fit, we show the results in Fig. 3, along with components of the bulk flow in Table 1, which are consistent with those found by LP. This confirms the results of other investigators who have found that the large magnitude of the bulk flow, pointed in a direction away from the principle axes of the error matrix $A_{ij}$, supports a large amplitude for the power spectrum, and places theories like CDM, or MDM, normalized to COBE, at approximately "three-sigma"—between the 95% and 99% probability contours. Note that there is a ridge of high likelihood that continues to very low $\Gamma$ and $\sigma_8$ (*i.e.*, with approximately constant CMB quadrupole $Q$). The bulk flow only tests the power spectrum on very large scales, causing this degeneracy between the parameters.

In Figure 4, we show the likelihood for the LP data based on a nine-point fit to both bulk flow and shear, as well as the components of the fit in Table 1. This fit spreads the peak of the likelihood function down to lower amplitudes—it is not the very smallest scales which are forcing the high-amplitude favored by the bulk flow, but the intermediate scales probed by the shear. (This is analagous to the high "Cosmic Mach Number" of Ostriker and Suto (1990), who compare the magnitude of large-scale flows to the velocity dispersion on small scales.) As discussed in Feldman & Watkins (1994), the bulk flow window function is peaked at scales $k^{-1} \gtrsim 100 h^{-1}$ Mpc, but has a non-negligeable tail extending to small scales. The shear window function (more precisely, the appropriate components of the combined bulk flow & shear window function) is peaked at $30 h^{-1}$ Mpc $\lesssim k^{-1} \lesssim 100 h^{-1}$ Mpc, also with a significant tail extending to smaller scales.

The maximum likelihood value of the shear has components of the same order as the diagonal terms in the error matrix $(A_{pq}^{-1})^{1/2}$; the data are unable to strongly distinguish from the shear expected for *no* clustering power, $P(k) = 0$. Thus, including this in the likelihood calculation pulls down the implied amplitude of the power spectrum—the detection of the absence of shear is not the absence of a detection of shear. Again, there is a ridge of high probability extending to very low $\Gamma$ and $\sigma_8$ indicative of the still-large scales probed by the shear.

Taking into account both the shear and bulk flow still favor a somewhat larger amplitude for the power spectrum at moderate values of $\Gamma$—even the small shear is not enough to compensate for the large bulk flow. Only when we consider all of the data, as above, is the amplitude consistent with $\sigma_8 \approx 1$; this implies that the LP flow is rather "cold" aside from the bulk flow.

In Figure 2, we show the likelihood for the RPK data for the bulk flow and shear. In this case, the bulk flow alone shows a mild detection of power (since there *is* a bulk flow!), consistent to about one sigma with the CMB dipole, but the shear + bulk flow together only give an upper limit, since the shear is small. Note that, because the sample is so sparse, the window functions are quite broad and have significant contributions on all scales, so the "shear" and "bulk flow" measured by RPK are quite different quantities than those measured by LP (Feldman & Watkins 1995).

## 4. Discussion

Several other analyses of the bulk flow calculated from the LP data set have been presented. LP themselves, along with Strauss *et al.* 1994, conclude that its large magnitude is a significant problem within the framework of all currently-favored theories of structure formation, when normalized to the COBE observation; they rule out these conventional theories (many of which correspond to points on our $\sigma_8$-$\Gamma$ plane) at 94-98% confidence limits using "frequentist" Monte Carlo techniques. Feldman & Watkins 1994 use some of the same techniques as Sec. 3. above, similarly ruling out these theories at the 95% or greater level. Our results are similar; when the bulk flow alone is considered, conventional theories are indeed disfavored at 95-99%; once more information is taken into account, however, these theories become significantly less unlikely.

These analyses have used "frequentist" statistics: they calculate some quantity for the LP data, such as the bulk flow or a $\chi^2$ that takes into account the spatial distribution of the data. Then, they calculate the probability that a particular model (*e.g.*, particular values of $\sigma_8$ and $\Gamma$) would produce a value of this statistic *as large or larger* than observed, given the known properties of the error distribution of the data. In some sense, these analyses integrate the likelihood over possible data sets. We perform a Bayesian analysis which uses the same probability distribution (our likelihood function) but integrates only over possible parameters as in Eq. 1. The two methods are not equivalent except in the case of linear models with uncorrelated errors, which this most definitely is not—the model, encoded in the quantity $R_{ij}$, appears only through the correlation matrix. In any case, when analyzing the same data (*i.e.*, the bulk flow), we stress that the two methods *do* agree—popular power spectra such as CDM are disfavored. Note that it is not immediately clear how to extend the frequentist analysis to the case with a small shear and large bulk flow—do we want the probability of observing a *smaller* or *larger* value of some suitable $\chi^2$?

Methodological differences aside, we agree with these analyses in the following sense: on the largest scales probed by the LP data, the perturbation amplitude appears to be quite large, as implied by the large bulk flow. On the intermediate scales probed by the shear, however, the amplitude is somewhat smaller—most of the components of the shear tensor are within "one sigma" of a purely poisson velocity field (*i.e.*, they are of comparable magnitude to the diagonal components of the error matrix $A_{pq}^{-1}$ of Sec. 3.). In order to fit best on all scales at once, we are forced to quite reasonable values of the amplitude of the power spectrum. Note, however, that this does not address the possibility that this particular parametrization of the power spectrum is inadequate to describe the data well.

For comparison, we observe a bulk flow $U = 842 \pm 398$ km/s and an rms shear $p = \langle p_{ij} p_{ij} \rangle = 12 \pm 10 h$ km/s/Mpc, where the large errors come from the sparse sampling of LP's data (*i.e.*, the large magnitude of the error part of the correlation matrix, $R_{mne}^{(\ )}$). Standard CDM ($\sigma_8 = 1$, $\Gamma = 0.5$), sampled with LP's window function, predicts $U \simeq 568$ km/s and $p \simeq 14$ km/s/Mpc. A power spectrum with $\sigma_8 = 2$, $\Gamma = 0.5$ that reproduces the observed bulk flow predicts a somewhat larger and non-negligeable $p \simeq 19$ km/s/Mpc. (None of these results have been corrected for LP's "error bias," and the errors all assume a diagonal correlation matrix between the points; for this reason we do not quote similar numbers for the RPK sample where the off-diagonal correlations are non-negligeable.)

Thus far, we have assumed that the errors on the line-of-sight velocities are simply diagonal; there are no correlations between them. This neglects the fact that the $L_m - \alpha$ relation is calibrated from the data itself, resulting in correlations between the line-of-sight velocities, since the standard candle luminosities depend on the whole data set. Moreover, the relation is calibrated assuming that the velocity field is well-modelled by a bulk flow superposed on essentially uncorrelated velocities (or uncorellated luminosity flucutations, depending on the details of the analysis as in Colless (1995)). This could potentially bias the reconstruction of any velocities other than the bulk flow (conversely, the presence of strong motions on smaller scales would bias the bulk flow result). As mentioned above, we have allowed the $L_m - \alpha$ relation to vary, assuming different bulk flow models, and the likelihood contours do not change substantially. We also note that the size of the scatter in the $L_m - \alpha$ relation is not an issue, even though it may in fact dominate over the actual velocities, as long as the distribution of the errors is well-understood.

Ideally, we should work with the observed $L_m$ and $\alpha$ observations directly and perform our own calibration, marginalizing over unknown nuisance parameters—like the paramaters of the $L_m - \alpha$ relation itself, which are not well-determined. (We pause here to note that "nuisance parameter" is a technical term, referring to parameters involving physics in which we are not presently interested.)

For the RPK data, the situation is less severe; they calibrate their distance indicator with a separate sample of galaxies, so these biases should not occur, although the parameters of the distance indicator relationship should in principle be marginalized over in this case as well. In any case, this is a statement about the technical ease of adapting the procedure of RPK to this context; ideally, all of the available information should be used in both cases.

Finally, how, if at all, do we reconcile the LP and RPK observations? With a sample of 13 galaxies, RPK observe a flow which they claim has nearly "converged" to the CMB dipole at the 7000 km/s depth of the survey, whereas LP find a strong bulk flow approximately 90° away from the CMB dipole. The two surveys seem to be probing quite different velocity fields, not surprising due to their very different effective window functions. Obviously, the discrepancy between this result and that of LP needs to be better understood: if it is not merely a rare statistical fluctuation, we may trace it back to some un-accounted for systematic bias in one of the distance indicators.

The authors would like to thank Dick Bond and Michael Strauss for valuable discussion and especially thank Marc Postman and Tod Lauer for providing their data as well as insight into its use. We also wish to thank Matthew Colless for providing the fits to his BCG distance indicator relation.

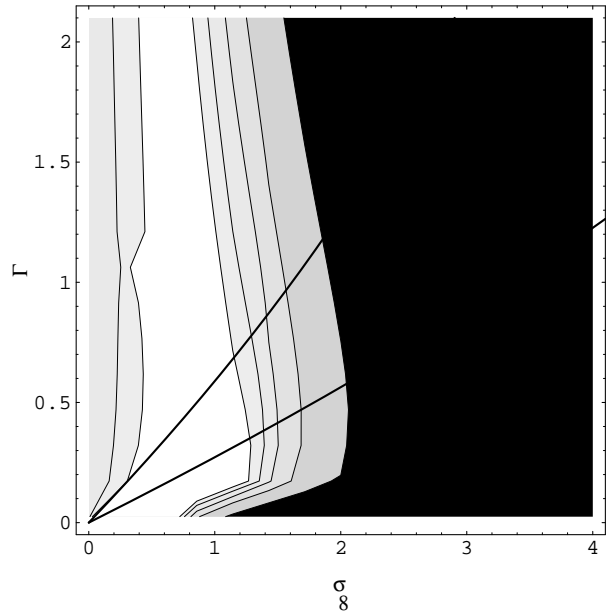

Fig. 1.— Full Likelihood as a function of power spectrum parameters $\sigma_8$ and $\Gamma$ for the clumped LP data. Contours of constant likelihood are plotted (see text for likelihood ratios relative to maximum). For uniform priors over the plotted area, the contours enclose 50%, 68%, 90%, 95% and 99% of the accumulated probability. Also shown are 68% confidence intervals for the COBE quadrupole (solid lines originating at the origin). We have assumed $\Omega = 1$; otherwise the horizontal axis should be replaced by the combination $f(\Omega)\sigma_8 = \Omega^{0.6}\sigma_8 = \beta$.

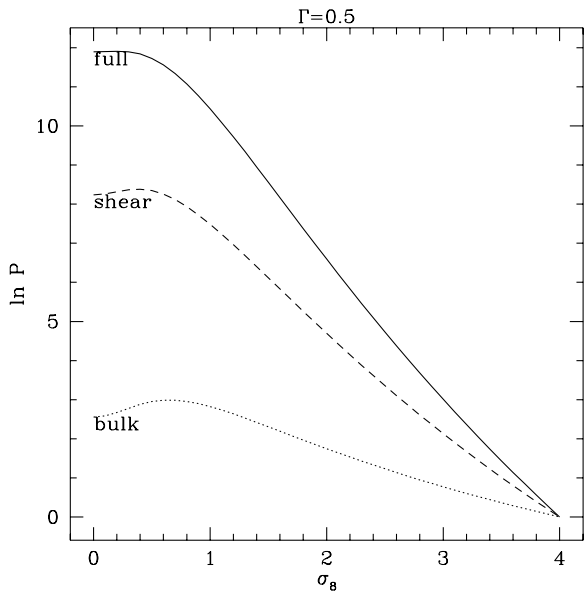

Fig. 2.— Likelihood function for RPK data. The three curves correspond to the full likelihood, the bulk flow + shear and the bulk flow alone. Because they use progressively less information, the full likelihood is the most restrictive, followed by the bulk flow + shear and the bulk flow alone.

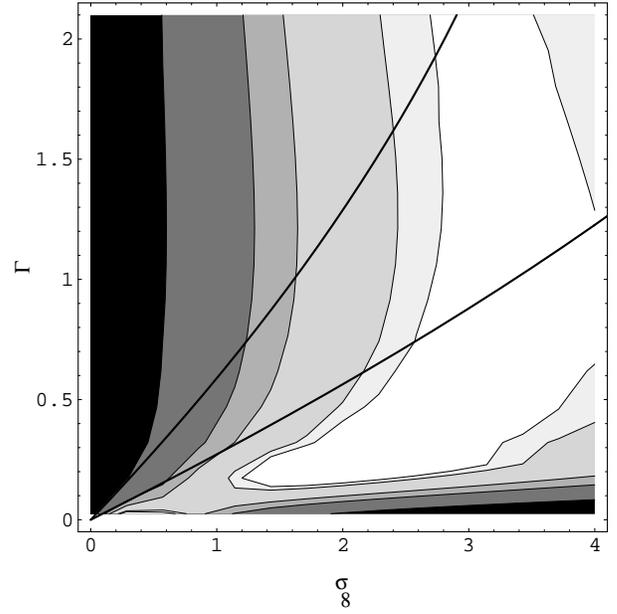

Fig. 3.— Bulk Flow Likelihood for the unclumped LP data, reduced to the 3 points of the bulk flow velocity. Contours of constant likelihood are shown, enclosing the same integrated probability as in Fig. 1.

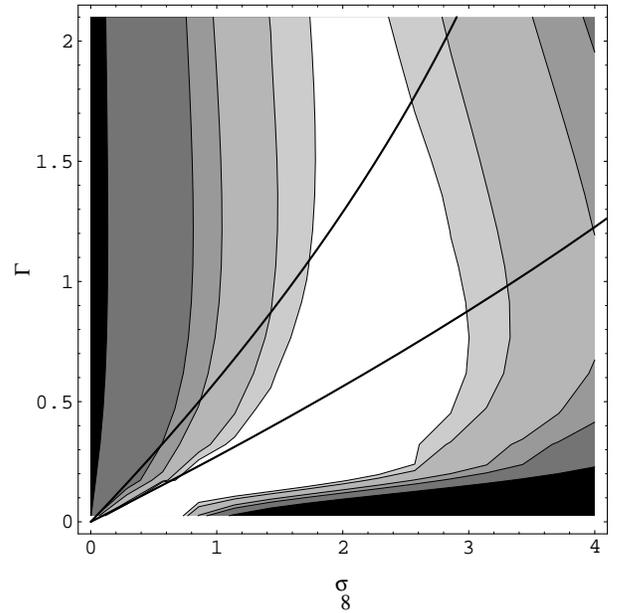

Fig. 4.— Bulk Flow + Shear Likelihood for the unclumped LP data, reduced to the 9 points of the bulk velocity and gradient. Contours of constant likelihood are shown, enclosing the same integrated probability as in Fig. 1.

|  | Bulk Flow [km/s] | | | Shear [$h$ km/s/Mpc] | | | | | |
| --- | --- | --- | --- | --- | --- | --- | --- | --- | --- |
|  | $U_x$ | $U_y$ | $U_z$ | $p_{xx}$ | $p_{xy}$ | $p_{yy}$ | $p_{xz}$ | $p_{yz}$ | $p_{zz}$ |
| Bulk | 530 | 80.0 | 649 | | | | | | |
| $\pm$ | 301 | 327 | 232 | | | | | | |
| Bulk & Shear | 484 | -216 | 644 | -6.96 | -5.18 | -1.48 | -2.49 | 2.76 | -1.22 |
| $\pm$ | 312 | 364 | 241 | 6.10 | 4.36 | 7.17 | 2.88 | 3.48 | 3.71 |

Table 1: Components of bulk flow (relative to the CMB) and shear for the LP dataset. The rows labeled "$\pm$" refer to the diagonal components of the error matrix $\sqrt{A_{pq}^{-1}}$.